\documentclass[preprint,12pt]{elsarticle}

\usepackage{amssymb}
\newcommand{\tool}[1]{\emph{DeVAIC}} 
\newcommand{\exbard}[1]{Gemini} 
\newcommand{\exbing}[1]{Copilot} 
\usepackage{comment} 
\usepackage[dvipsnames]{xcolor} 
\usepackage{amsmath} 
\usepackage{listings} 
\usepackage{array}
\usepackage{xcolor}
\usepackage{booktabs}
\usepackage{multirow,bigstrut}
\usepackage{tabu}
\newlength\MAX 
\newcommand*\Chart[1]{#1~\rlap{\textcolor{black!20}{\rule{\MAX}{2ex}}}\rule{#1\MAX}{2ex}}
\usepackage[caption=false, font=footnotesize]{subfig}
\usepackage{hyperref}
\usepackage{nameref}
\usepackage{todonotes}
\newcommand{\pietro}[2][]{\todo[color=red!50, , inline, #1]{\textbf{Pietro:} #2}}
\newcommand{\roby}[2][]{\todo[color=violet!40, , inline, #1]{\textbf{Roby:} #2}}

\journal{Information and Software Technology}

\begin{document}

\begin{frontmatter}


\title{\tool{}: A Tool for Security Assessment of AI-generated Code}
\author{Domenico Cotroneo}
\ead{cotroneo@unina.it}
\author{Roberta De Luca\corref{cor1}}
\ead{roberta.deluca2@unina.it}
\author{Pietro Liguori}
\ead{pietro.liguori@unina.it}
\cortext[cor1]{Corresponding author}
\affiliation{organization={University of Naples Federico II},
             city={Naples},
             postcode={80125},
             country={Italy}}


\begin{abstract}
\noindent
\textbf{Context}: AI code generators are revolutionizing code writing and software development, but their training on large datasets, including potentially untrusted source code, raises security concerns. Furthermore, these generators can produce incomplete code snippets that are challenging to evaluate using current solutions.\\
\textbf{Objective}: This research work introduces \tool{} (Detection of Vulnerabilities in AI-generated Code), a tool to evaluate the security of AI-generated Python code, which overcomes the challenge of examining incomplete code.\\
\textbf{Method}: We followed a methodological approach that involved gathering vulnerable samples, extracting implementation patterns, and creating regular expressions to develop the proposed tool. The implementation of \tool{} includes a set of detection rules based on regular expressions that cover $35$ Common Weakness Enumerations (CWEs) falling under the OWASP Top $10$ vulnerability categories.\\
\textbf{Results}: We utilized four popular AI models to generate Python code, which we then used as a foundation to evaluate the effectiveness of our tool. \tool{} demonstrated a statistically significant difference in its ability to detect security vulnerabilities compared to the state-of-the-art solutions, showing an $F_1$ Score and Accuracy of $94\%$ while maintaining a low computational cost of $0.14$ seconds per code snippet, on average.\\
\textbf{Conclusions}: The proposed tool provides a lightweight and efficient solution for vulnerability detection even on incomplete code.
\end{abstract}



\begin{keyword}
Static code analysis \sep Vulnerability detection \sep AI-code generators \sep Python



\end{keyword}

\end{frontmatter}


\section{Introduction}
\label{sec:introduction}
We live in an era where AI-code generators are revolutionizing the process of writing code and software development.
AI-powered solutions like GitHub Copilot~\cite{copilot}, OpenAI ChatGPT~\cite{chatGPT}, Google \exbard{}~\cite{bard}, and Microsoft \exbing{}~\cite{bing} have shown their ability to translate into programming code what users request with natural language (NL) descriptions (e.g., English language).

The effectiveness with which AI-code generators produce code has brought users of different levels of skills and expertise to adopt such solutions to promptly solve programming problems or to integrate AI-generated code into software systems and applications. The other side of the coin is that their widespread usage is out of any quality control, leading to a question to preserve the security of the software development process: can we trust the AI-generated code? \textit{Ergo}, from a software security perspective, is the code generated by AI secure and free of software vulnerabilities?

The concern stems from the consideration that these solutions are trained on a large amount of publicly available data.
For instance, GitHub Copilot utilizes training data that consists of an extensive amount of source code, with billions of lines collected from publicly accessible sources, including code from GitHub's public repositories~\cite{github_blog}.
Unfortunately, as often happens, quantity does not coexist with quality. 
In fact, the huge amount of data used to train the AI-code generators may include deprecated functions and libraries, which may lead to the exploitation of vulnerabilities when adopted, or it could intentionally have buggy or insecure code used to \textit{poison} the code generators during the training phase~\cite{improta2023poisoning,li2023poison,hussain2023trojanedcm,hamer2024just}.

This issue is not unexpected at all. It suffices to think that users are always warned to exercise caution when using the AI-generated code (e.g., ``\textit{The users of Copilot are responsible for ensuring the security and quality of their code}''~\cite{warn_copilot}, ``\textit{Use the code with caution}''~\cite{bard_blog}, ``\textit{This code is generated by artificial intelligence. Review and use carefully}''~\cite{bing_blog}).

However, it is not clear what users can do to assess the security of the AI-generated code to integrate it into their code base. 
\textit{Manual analysis}, the go-to method for security experts, becomes unfeasible due to the volume and rate deployment of AI-generated code~\cite{liguori2021evil,liguori2022can}.
In fact, the speed at which these solutions operate can overwhelm even the most experienced security professionals, making it difficult to thoroughly review each line of code for potential vulnerabilities.

To perform code vulnerability detection, state-of-the-art solutions provide several \textit{static analysis tools}. 
Such tools usually parse the Abstract Syntax Tree (AST) of the code~\cite{ma2022code,bandit,pyt,codeql}, a hierarchical representation of its structure, and require complete programs to check whether the code contains software vulnerabilities. 
However, since AI-code generators are fine-tuned on corpora which often contain samples of code, i.e., \textit{code snippets}~\cite{humaneval,django,liguori-etal-2021-shellcode,conala}, they often do not produce complete programs~\cite{liguori2022can,zhong2023study}, making infeasible the application of these tools in this context. 


A different solution is represented by tools implementing pattern-matching approaches to detect software vulnerabilities~\cite {mastjik2015comparison,patel2015implementation,walenstein2007exploiting}. These tools require users to set up ad-hoc configuration files to specify the vulnerabilities they wish to detect via matching patterns~\cite{mordahl2023automatic,nadi2014mining}, hereby requiring a non-trivial manual effort and limiting their adoption in practice.

Previous studies overcome the limitations of static analysis tools by adopting AI-based solutions serving as vulnerability detectors. These fully automated solutions do not require any human effort and can analyze incomplete programs, but they may be susceptible to a high rate of false alarms~\cite{chakraborty2021deep,chen2023diversevul,li2021vulnerability,li2018vuldeepecker}. 
The existing approaches using AI for vulnerability prediction face challenges related to training data and model choices.
Furthermore, it has been shown that AI models struggle to understand the complexity of code structures and identify security issues~\cite{fang2023large,al2023chatgpt,cheshkov2023evaluation}.

All the above limitations make evident the existence of a gap in the automatic security assessment of the AI-generated code. Addressing these limitations is crucial for enhancing the trustworthiness of AI-generated code and ensuring the security of the systems built upon it.

This paper presents \textbf{\tool{}} (\textbf{\textit{De}}tection of \textbf{\textit{V}}ulnerabilities in \textbf{\textit{AI}}-generated \textbf{\textit{C}}ode), a tool that performs static analysis of Python code by implementing a set of detection rules. 
More precisely, the tool uses a set of regular expressions that cover $35$ Common Weakness Enumeration (CWE), a community-developed list of software and hardware weakness types, to detect vulnerabilities in Python code, one of the most used programming languages~\cite{octoverse,meyerovich2013empirical,statista}.
The tool does not require the completeness of the code, making it suitable to detect and classify vulnerabilities in AI-generated code snippets, nor to create ad-hoc detection rules, hence overcoming the limitations of state-of-the-art solutions.

We used \tool{} to detect vulnerabilities in the code generated by four well-known public AI-code generators starting from NL prompts. Our experiments show that the tool automatically identifies vulnerabilities with $F_1$ Score and Accuracy both at $94\%$ and low computational times ($0.14$ seconds for code snippet, on average). 
Also, we show that \tool{} exhibits performance superior to the state-of-the-art solutions, i.e. CodeQL~\cite{codeql}, Bandit~\cite{bandit}, PyT~\cite{pyt} and Semgrep~\cite{semgrep}, and the models ChatGPT-$3.5$~\cite{chatGPT}, ChatGPT-$4$, and Claude-$3.5$-Sonnet~\cite{claude35sonnet}, that are widely used to perform vulnerability detection of the code~\cite{pearce2022asleep,dunlap2023finding,bakhshandeh2023using,kathikar2023assessing}.

In the following, Section~\ref{sec:related} discusses related work; Section~\ref{sec:motivating} introduces a motivating example; Section~\ref{sec:methodology} presents \tool{}; Section~\ref{sec:evaluation} describes the results; Section~\ref{sec:threats} discusses the threats to validity; Section~\ref{sec:conclusion} concludes the paper.

\section{Related Work}
\label{sec:related}
Static analysis is one of the most commonly used techniques to detect vulnerabilities in the code~\cite{chess2004static,li2017static,pan2020systematic,goseva2015capability}. 

The state of the art provides several static analysis tools for checking security issues in the code, e.g., Python-specific tools like Bandit~\cite{bandit} and PyT~\cite{pyt}, or multilanguage ones such as Semgrep~\cite{semgrep} and CodeQL~\cite{codeql}, which are widely used to detect vulnerabilities within the code~\cite{pearce2022asleep,dunlap2023finding,bakhshandeh2023using,kathikar2023assessing,ruohonen2021large,kapustin2023static,peng2019python,lyons2021meta,bandara2020fix,gobbi2023poster,cherry2022static}.

Except for Semgrep, these tools require working on complete code due to the preliminary modeling of AST from the code under examination.
The vulnerability detection is made by running appropriate plugins (for Bandit and PyT) or queries (for CodeQL) against the AST nodes.
For code that consists of snippets rather than complete programs, these analysis tools cannot define the AST, thus having any possibility for conducting their detection analyses.
Semgrep~\cite{semgrep} is a static analysis tool that uses a pattern-matching approach and that does not require the AST modeling of code before running the detection rules. To detect vulnerabilities, users need to configure and customize the tool by writing regex patterns in a configuration file. 
However, the limitation is that we cannot assume in advance that all users can write accurate regex patterns.
Moreover, this approach could polarize vulnerability hunting, focusing on those the user believes to find, potentially overlooking others that are effectively present~\cite{mordahl2023automatic,nadi2014mining}. 
For this reason, as often happens, this type of solution offers a set of rules publicly available for the scanning of the code under examination~\cite{semgrep_registry}.


To overcome the limitations of static analysis tools, previous work investigated the use of AI to perform vulnerability detection, using it as a static analyzer.
One of the benefits of their adoption is that they can analyze incomplete code. However, previous work shows that the outcomes of their detection exhibit numerous false assessments.

For instance, Chen \textit{et al.}~\cite{chen2023diversevul} released a new vulnerable source code dataset to assess state-of-the-art deep learning methods in detecting vulnerabilities. 
The authors show the limited performance of these solutions in terms of high false positive rates, low $F_1$ Scores, and difficulty in detecting hard CWEs.
Ullah \textit{et al.}~\cite{ullah2024llms} used 17 prompt engineering techniques to test 8 different LLMs in vulnerability detection. Among the 228 code scenarios employed in the experiments, the LLMs frequently mistakenly identify patched examples as vulnerable, causing an elevated number of false positives.
Similarly, Purba \textit{et al.}~\cite{purba2023software} encountered a high rate of false positives when using language models like GPT-3.5, CodeGen, and GPT-4 to classify a snippet of code (i.e., a segment of code) as vulnerable or not vulnerable.
Fang \textit{et al.}~\cite{fang2023large} crafted a dataset that pairs real-world code with obfuscated versions, which they used as input for large language models (LLMs) to evaluate their ability to analyze input code samples and test if LLMs can be employed for defensive analysis tasks. 
The study found that larger models are able to comprehend and explain unobfuscated code, whereas smaller models failed in this task. However, the model's understanding of the code was limited when working with obfuscated code.
Al-Hawawreh \textit{et al.}~\cite{al2023chatgpt} and Cheshkov \textit{et al.}~\cite{cheshkov2023evaluation} evaluated the performance of the ChatGPT for detecting vulnerabilities in code, finding that ChatGPT's results still needed careful monitoring.
Khoury \textit{et al.}~\cite{khoury2023secure} explored how safe is the code generated by ChatGPT, using this model for both code generation and assessment. Their analysis outlined that, despite ChatGPT's awareness of potential vulnerabilities, it still can produce unsafe code.
Instead of using AI-based solutions, Sandoval \textit{et al.}~\cite{sandoval2023lost}~\cite{sandoval2022security} assessed the security of AI-generated code by performing a manual code analysis. Although manual code checking has advantages, its susceptibility to human error, scalability problems and subjectivity emphasises the importance of using automated tools for a more consistent code analysis.

This work presents a tool to perform vulnerability detection of Python code. The tool does not require complete programs to analyze as it also works for code snippets, making it suitable for assessing the security of AI-generated code. Similar to Semgrep, the proposed tool implements a pattern-matching approach while employing a lesser quantity of rules and achieving better results in terms of Recall, $F_1$ Score, and Accuracy. Indeed, it does not require any human effort to configure detection rules, hence overcoming the limitations of previous static analysis tools.

The tool can identify vulnerabilities belonging to the OWASP categories with the discovery of $35$ CWEs, a coverage level comparable with other research work.
For instance, Hajipour \textit{et al.}~\cite{hajipour2023systematically} consider $13$ CWEs for the assessment of the security risks posed by code language models through an approximation of the inversion of the black-box models using few-shot prompting; Pearce \textit{et al.}~\cite{pearce2022asleep}, moreover, refer to the MITRE’s Top $25$ CWE list to assess the code generated by Copilot.
CodeQL provides queries to assess the security of Python code with a coverage of $44$ CWEs~\cite{codeql_coverage}.  
Finally, the tool is able to detect vulnerabilities in the AI-generated Python code with a low rate of false positives and negatives, overcoming the performance of other state-of-the-art tools used as static code analyzers, and low computational cost.

\section{Motivating Example}
\label{sec:motivating}
AI models for code generation have shown significant advancements, largely attributed to their intuitive, user-friendly interfaces. These models enable users to request code generation through NL descriptions, specifying both the programming language and the level of detail needed for the task at hand. For illustrative purposes, consider the following example: 

\begin{center}
\begin{lstlisting}[caption={Example of NL prompt.}, captionpos=b, basicstyle=\footnotesize]
Write a Python code that defines a YAML payload from a request, 
loads this payload into a dictionary called prod_and_qt
by using a function for loading.
\end{lstlisting}    
\end{center}

Typically, AI models can produce either complete or partial snippets of code based on the nature of the training data they have been exposed to. 
A concrete instance of AI-generated code for the aforementioned prompt might look like this:

\begin{center}
\begin{lstlisting}[language=Python, captionpos=b, caption={Example of code generated by AI models.}, basicstyle=\footnotesize]
def load_yaml(yaml_payload):
   prod_and_qt = yaml.load(yaml_payload)
   return prod_and_qt
   
yaml_payload = """
product:
 name: iPhone
 quantity: 10
"""

prod_and_qt = load_yaml(yaml_payload)
print(prod_and_qt)
\end{lstlisting} 
\end{center}

The code generated by the model is what is requested in the NL prompt.
However, the code is vulnerable due to the \texttt{yaml.load()} function used to process YAML data from \texttt{yaml\_payload}, which may contain untrusted content. 
Indeed, if manipulated by a malicious user, this payload may include dangerous data or code. 
The official \texttt{PyYAML} documentation~\cite{PyYAML_doc} advises against using the \texttt{yaml.load()} function because it interprets the YAML payload as Python code, potentially allowing the execution of malicious instructions if the code lacks proper validation, including calls to the \texttt{os.system} library, which can execute any command on the system.
The CWE associated with this vulnerability is \textit{CWE-502}, commonly known as \textit{Deserialization of Untrusted Data}, and related to the \textit{Software and Data Integrity Failures} category of OWASP's Top 10. 
A simple way to address this issue is reported by the official \texttt{PyYAML} documentation, which recommends using \texttt{yaml.safe\_load()} to read YAML from unreliable and untrusted sources~\cite{PyYAML_doc}.
The \texttt{yaml.safe\_load()} function is designed to limit the types of objects that can be loaded to standard ones (e.g., dictionaries, lists, strings, numbers, etc.), thus avoiding the execution of arbitrary code.

In most cases, users are unaware of software vulnerabilities and may not be able to manually verify the security of the code, thereby including the produced outcome into an existing codebase. 
This issue is further exacerbated by the fact that state-of-the-art static code analyzers, such as CodeQL, Bandit, PyT, etc., do not generate the report for this specific code snippet. In fact, the code reported above is \textit{incomplete} due to the lack of the \texttt{import} statement at the beginning of it, making these tools unable to perform the vulnerability detection, as explained in Section~\ref{sec:related}.
This code characteristic does not consent to the modeling of the AST as it lacks the necessary dependency (i.e., \texttt{import yaml}) for the invoked API (i.e., the \texttt{yaml.load()} function).  
Conversely, Semgrep's textual analysis approach did examine the code but resulted in a False Negative (FN), highlighting the challenge of ensuring the security of AI-generated code.

The generation of incomplete code is a well-known problem with AI code generators. Recent works are currently addressing the issue by using prompt engineering techniques. For example, in~\cite{tony2024prompting} authors used prompt-engineering techniques to stimulate the model to generate secure code in Python, but the model occasionally produced incomplete code. To overcome this issue, they employed an iterative code-generation process by concatenating the prompts with the incomplete output generated by the models until they obtained a complete code. In ~\cite{li2024approach}, the authors explicitly requested the model to generate complete code by using precise system prompts that explicitly require the completeness of the code. Moreover, they employed multiple iterations to guarantee the correctness of the generated code.
Although prompt engineering can be an effective way to reduce the incompleteness in the code generated by the models (e.g., by using explicit requests or performing multiple iterations), we believe that, in a typical scenario, users would request models to generate code without being aware of the potential to produce incomplete code.

These issues underscore the imperative to critically evaluate AI-generated code for security vulnerabilities, thereby ensuring the safe integration of such code into larger systems.

\section{\tool{} Workflow}
\label{sec:methodology}
To overcome the issues described in Section~\ref{sec:motivating}, we present \tool{}, a tool to detect vulnerabilities in Python code. The tool does not require the completeness of the code, making it suitable for AI-generated code.

The \tool{} tool works as a text scanner and does not require the AST modeling for the code analysis, employing regular expressions to identify vulnerable code patterns. 
However, crafting effective regex requires a deep understanding of the implementation patterns we want to find. 

\begin{figure}[t!]
    \centering
    \includegraphics[width=1\columnwidth, trim=0 0.7cm 0 0.7cm, clip=true]{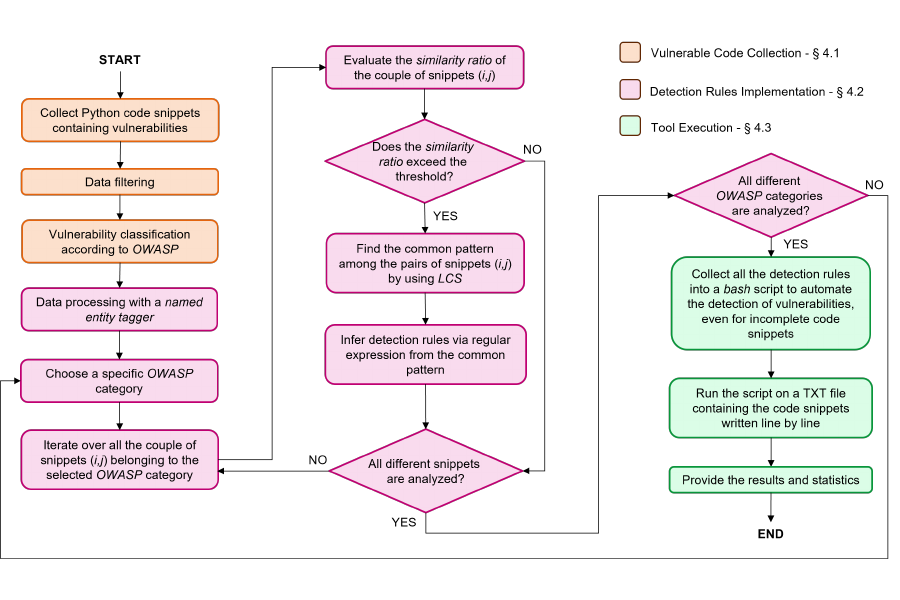}
    \caption{The \tool{} workflow.}
    \label{Fig:scheme}
\end{figure}

To this aim, we extensively reviewed the literature to collect datasets of unsafe Python code with the associated information of the implemented CWE\footnote{The Common Weakness Enumeration (CWE) is a list of common software and hardware weakness types published annually by the Massachusetts Institute of Technology Research and Engineering® (MITRE) Corporation.}, extracting only the snippets that implement the most recent and critical weaknesses identified in OWASP's Top $10$ Vulnerabilities\footnote{The Open Web Application Security Project® (OWASP) draws up the Top $10$ Application Security List every four years to define the $10$ most prevalent web application vulnerabilities.} report.
Since OWASP categories encompass various CWEs that share the same vulnerability typology, we employ these categories to cluster snippets together for further detailed analysis (see \S{}~\ref{snippetColl}).

Using a \textit{named entity tagger}, we standardize snippets by replacing variable names, and input and output parameters of functions to reduce the variability of the code.  
Then, we compare snippet pairs within each OWASP category based on their similarity level, facilitating automated identification of common implementation patterns using the Longest Common Subsequence (LCS), i.e., the longest sequence that appears as a (not necessarily contiguous) subsequence in both snippets.
By using the identified patterns, we infer regex-based \textit{detection rules} able to identify vulnerabilities with similar patterns and assess their corresponding vulnerability types (see \S{}~\ref{detrules}).

Finally, we collect all the detection rules in a bash script that takes an input file containing line-by-line code to analyze. After scanning the file, the tool generates a report of the detection, including the number of snippets identified as vulnerable and their category according to the OWASP (see \S{}~\ref{tool_presentation}).

\figurename{}~\ref{Fig:scheme} shows a detailed flowchart outlining the steps through which we developed the tool.
In the rest of this Section, we detail the steps of the workflow.

\subsection{Vulnerable Code Collection}
\label{snippetColl}

In our pursuit to collect vulnerable snippets, we performed a comprehensive study of the literature. State-of-the-art provides numerous corpora containing vulnerable code, some aimed at poisoning models to induce them to generate vulnerabilities, while others serve to evaluate AI models as vulnerability detectors~\cite{cotroneo2023vulnerabilities,chen2023diversevul}. 

Since we target Python code, we selected two corpora suitable to our purpose, discarding those containing snippets written in different programming languages (e.g., DiverseVuln~\cite{chen2023diversevul}, Big-Vul~\cite{bigvul,fan2020ac}, SARD~\cite{sard}): 
\begin{enumerate}
    \item \textit{SecurityEval}~\cite{seceval}: It is a dataset built to evaluate the security of code generation models and to compare the performance of some state-of-the-art static analysis tools, such as Bandit and CodeQL~\cite{siddiq2022seceval}. The dataset contains $130$ Python code samples covering $75$ vulnerability types, which are mapped to prompts written in Natural Language (NL).
    \item \textit{Copilot CWE Scenarios Dataset}~\cite{datasetCWE}: It encompasses the source code utilized to craft NL prompts for generating code with Large Language Models (LLMs)~\cite{llmseceval,llmseceval2023}. It covers code associated with $89$ distinct vulnerable scenarios.
\end{enumerate}

\begin{table}[h!]
\centering
\caption{List of $35$ selected CWEs. In \textcolor{blue}{\textbf{blue}} we indicate the CWEs belonging to at least one of the MITRE's top $25$ of the last three years. We use ``-'' to indicate the absence of a specific CWE in the relative top $40$.}
\label{tab:cwe_details_owasp}
\scriptsize 
\begin{tabular}{ 
>{\raggedright\arraybackslash}m{7cm} | 
>{\centering\arraybackslash}m{1.5cm} | 
>{\centering\arraybackslash}m{1cm} |
>{\centering\arraybackslash}m{1cm} |
>{\centering\arraybackslash}m{1cm}}
\toprule
\textbf{OWASP 2021} & \textbf{CWE} & \textbf{Rank MITRE 2021} & \textbf{Rank MITRE 2022} & \textbf{Rank MITRE 2023} \\
\toprule
\multirow{4}{6.5cm}{\textit{Broken Access Control}}
& \textcolor{blue}{CWE-022} & 8 & 8 & 8 \\ 
& CWE-377 & - & - & - \\ 
& CWE-425 & - & - & - \\ 
& CWE-601 & 37 & 35 & 32 \\
\midrule
\multirow{9}{6.5cm}{\textit{Cryptographic Failures}}
& CWE-319 & 35 & 39 & - \\ 
& CWE-321 & - & - & - \\ 
& CWE-326 & - & - & - \\ 
& CWE-327 & - & - & - \\ 
& CWE-329 & - & - & - \\ 
& CWE-330 & - & - & - \\ 
& CWE-347 & - & - & - \\ 
& CWE-759 & - & - & - \\ 
& CWE-760 & - & - & - \\
\midrule
\multirow{2}{6.5cm}{\textit{Identification and Authentication Failures}}
& CWE-295 & 26 & 26 & 34   \\
& CWE-384 & - & - & -   \\ 
\midrule
\multirow{10}{6.5cm}{\textit{Injection}}
& \textcolor{blue}{CWE-020} & 4 & 4 & 6   \\
& \textcolor{blue}{CWE-078} & 5 & 6 & 5   \\
& \textcolor{blue}{CWE-079} & 2 & 2 & 2   \\
& CWE-080 & - & - & -   \\
& CWE-090 & - & - & -   \\
& \textcolor{blue}{CWE-094} & 28 & 25 & 23   \\
& CWE-095 & - & - & -   \\
& CWE-096 & - & - & -   \\
& CWE-099 & - & - & -   \\
& CWE-113 & - & - & -   \\
& CWE-116 & - & - & -   \\
& CWE-643 & - & - & - \\ 
& CWE-1236 & - & - & -   \\
\midrule
\multirow{3}{6.5cm}{\textit{Insecure Design}}
& CWE-209 & - & - & - \\
& \textcolor{blue}{CWE-269} & 29 & 29 & 22 \\ 
& \textcolor{blue}{CWE-434} & 10 & 10 & 10 \\ 
\midrule
\textit{Security Logging and Monitoring Failures} (SLMF)
& CWE-117 & - & - & - \\ 
\midrule
\textit{Security Misconfiguration}
& \textcolor{blue}{CWE-611} & 23 & 24 & 28 \\ 
\midrule
\textit{Server-Side Request Forgery} (SSRF)
& \textcolor{blue}{CWE-918} & 24 & 21 & 19\\
\midrule
\textit{Software and Data Integrity Failures} (SDIF)
& \textcolor{blue}{CWE-502} & 13 & 12 & 15 \\ 
\bottomrule
\end{tabular}
\end{table}

To develop detection rules that address the most recent and critical weaknesses, we extracted from these corpora only the vulnerable code snippets that implement the CWEs present in the most recent OWASP Top $10$ Vulnerabilities report from 2021~\cite{owasp_2021,honkaranta2021towards}.
Overall, we selected $240$ vulnerable snippets that implement $35$ CWEs related to $9$ OWASP categories\footnote{MITRE itself is the authority responsible for establishing the correlation between CWEs and the OWASP Top $10$}. 
Among the selected CWEs, $10$ of them can be found in at least one of the MITRE's Top $25$ lists from the past three years~\cite{top40_2021,top40_2022,top25_2023}. 
The use of $35$ CWEs is consistent with other detection setups used in previous research studies~\cite{hajipour2023systematically,pearce2022asleep}. 
\tablename{}~\ref{tab:cwe_details_owasp} shows the selected CWEs with their OWASP category.

\subsection{Detection Rules Implementation}
\label{detrules}

To define detection rules, we search for common patterns among the selected code snippets based on the assumption that similar vulnerabilities manifest in similar ways~\cite{thomas2020learning,honkaranta2021towards,mastjik2015comparison}.
Pattern-matching approaches such as Longest Common Subsequence (LCS), N-gram and others, have been widely used to identify the same family malware~\cite{mastjik2015comparison,walenstein2007exploiting,patel2015implementation,zhang2022example}. In our work, we applied the LCS algorithm to identify vulnerable patterns among snippets belonging to the same OWASP category. 

As an example, consider the pair of code snippets shown in \tablename{}~\ref{tab:snippets} - ``Original Snippet'' column, which belongs to the Injection OWASP category (i.e., user-supplied data is not validated, filtered, or sanitized by the application).
Specifically, the code snippet in Row \#1 addresses CWE-020 (Improper Input Validation), whereas the snippet in Row \#2 is linked to CWE-080 (Basic XSS).
Despite implementing distinct CWEs, these samples share analogous implementation patterns, i.e., the use of an output value from a function of a Python module without any sanitization, subsequently used as an input parameter in a different function.

\begin{table}[!h]
\centering
\caption{Examples of standardization for code snippets. The common longest subsequences of original snippets and the standardized snippets are in \textcolor{blue}{\textbf{blue}}.}
\label{tab:snippets}
\footnotesize
\begin{tabular}{
>{\centering\arraybackslash}m{1cm} | 
>{\raggedright\arraybackslash}m{5.75cm} |
>{\raggedright\arraybackslash}m{5.75cm} }
\toprule
 & \textbf{Original Snippet} & \textbf{Standardized Snippet}\\
\toprule
S1 & \texttt{def temperature\_for\_locatio\textcolor{blue}{\textbf{n}}(): file\_y\textcolor{blue}{\textbf{am}}l \textcolor{blue}{\textbf{=}} \textcolor{blue}{\textbf{request.args.get(}}"fil\textcolor{blue}{\textbf{e}}"\textcolor{blue}{\textbf{,}} None\textcolor{blue}{\textbf{)}} j\textcolor{blue}{\textbf{son}}\_payload \textcolor{blue}{\textbf{=}} ya\textcolor{blue}{\textbf{m}}l.lo\textcolor{blue}{\textbf{a}}d(fil\textcolor{blue}{\textbf{e}}\_y\textcolor{blue}{\textbf{am}}l\textcolor{blue}{\textbf{)}}} &
\texttt{def temperature\_for\_location(): \textcolor{blue}{\textbf{var0 = request.args.get(var1, var2) var3 =}} y\textcolor{blue}{\textbf{a}}ml.l\textcolor{blue}{\textbf{o}}ad\textcolor{blue}{\textbf{(var0)}}}\\
\midrule
S2 & \texttt{\textcolor{blue}{\textbf{nam}}e \textcolor{blue}{\textbf{=}} \textcolor{blue}{\textbf{request.args.get(}}`nam\textcolor{blue}{\textbf{e}}'\textcolor{blue}{,} `'\textcolor{blue}{\textbf{)}} re\textcolor{blue}{\textbf{s}}p\textcolor{blue}{\textbf{on}}se \textcolor{blue}{\textbf{=}} \textcolor{blue}{\textbf{ma}}k\textcolor{blue}{\textbf{e}}\_response(n\textcolor{blue}{\textbf{am}}e\textcolor{blue}{\textbf{)}} return response} &
\texttt{\textcolor{blue}{\textbf{var0 = request.args.get(var1, var2) var3 =}} m\textcolor{blue}{\textbf{a}}ke\_resp\textcolor{blue}{\textbf{o}}nse\textcolor{blue}{\textbf{(var0)}} return var3}\\
\bottomrule
\end{tabular}
\end{table}

Therefore, to search for similar code snippets, we first check the similarity for each pair of snippets grouped in the same OWASP category. 
To this aim, we use \texttt{SequenceMatcher}, a class of the Python module \texttt{difflib}, that computes the Longest Common Subsequence (LCS) with no junk elements, i.e., tokens that are not considered in the matching (e.g., the newline character \texttt{$\backslash n$}), among different code snippets.
First, we compute the \textit{similarity ratio} between two code snippets, ranging from $0$ (total mismatching) to $1$ (perfect matching).
Based on empirical observations, we focus exclusively on pairs exhibiting a similarity ratio exceeding $50\%$~\cite{rao2018efficient,rao2018characteristic,rao2018partial}. 
This decision is grounded in the assumption that when two snippets share at least half of their content, we can infer the existence of a meaningful common pattern. 
Then, we find common patterns among snippets that meet the similarity threshold within each OWASP category by computing the LCS~\cite{ma2017vurle,mastjik2015comparison}.

To support the LCS in finding common patterns,
we \textit{standardize} all the snippets for each OWASP category, i.e., we reduce the randomness of the code snippets, by using a  \textit{named entity tagger}, which returns a dictionary of standardizable tokens for the input and output parameters of functions, extracted through regular expressions. 
We replace the selected tokens in every intent with ``\textit{var\#}'', where \textit{\#} denotes a number from $0$ to $|l|$, and $|l|$ is the number of tokens to standardize. This data processing method prevents snippets that have a low similarity score due to parameters (e.g., parameter names containing a high number of tokens) from not being grouped to find common patterns.

\tablename{}~\ref{tab:snippets} - ``Standardized Snippet'' column shows the standardization process performed on the snippets, where the parameters of the original code snippets are replaced (e.g., the input parameter ``file'' is replaced with \texttt{var1} in the code snippet of row $\# 1$). 
Due to the standardization, the similarity ratio increases from $\sim37\%$ (obtained on the original snippets) to $\sim63\%$.
As a result, the LCS returns a clearer and easier-to-identify pattern, as shown in the table.

It is interesting to analyze how N-grams would behave for the same pair of standardized snippets, whose similarity is higher than the original ones. To the best of our knowledge, N-grams can be used with two granularities, i.e., character N-grams and word (token) N-grams~\cite{chernis2018machine,khreich2017anomaly,lecluze2013granularity,mcnamee2004character}, and the typical values of N for the common pattern extraction are 4 or 6~\cite{khreich2017anomaly,lecluze2013granularity}. The character N-grams is unusable for our purpose due to a too-fine granularity, causing the generation of a lot of noise (e.g., referring to \tablename{}~\ref{tab:snippets} - ``Standardized Snippet'' column, we had the common patterns ['\texttt{var0}', '\texttt{ar0\_}', '\texttt{r0\_=}', '\texttt{0\_= }', '\texttt{\_= r}', '\texttt{= re}', '\texttt{req}', etc.] for character 4-grams, and ['\texttt{var0\_= }', '\texttt{ar0\_= r}', '\texttt{r0\_= re}', '\texttt{0\_= req}', etc.] for character 6-grams). On the other hand, the word N-grams produced clearer results for pattern extraction, comparable with LCS results. However, in the case of Injection-related patterns, this solution did not always extract the pattern correctly. Referring once again to the standardized snippets in \tablename{}~\ref{tab:snippets}, we obtained the common patterns [('\texttt{var0}', '\texttt{=}', '\texttt{request.args.get(var1,}', '\texttt{var2)}'), ('\texttt{=}', '\texttt{request.args.get(var1,}',\texttt{ 'var2)}', '\texttt{var3}'), ('\texttt{request.args.get(var1,}', '\texttt{var2)}', '\texttt{var3}', '\texttt{=}'] for word 4-grams, [('\texttt{var0}', '\texttt{=}', '\texttt{request.args.get(var1,}', '\texttt{var2)}', '\texttt{var3}', '\texttt{=}')] for word 6-grams. The pattern resulting from the word 6-grams is near to the one we extracted from LCS, as shown in blue in \tablename{}~\ref{tab:snippets} - ``Standardized Snippet''. However, with the word 6-grams we lost the information about the use of the variable \texttt{var0}, which contains the output of the \texttt{request.args.get()} function. In the snippets S1 and S2, \texttt{var0} is passed as a parameter in another function, and, if not sanitized, can introduce flaws in the code. The use of \texttt{var0} as a parameter is intercepted by LCS (i.e., \texttt{var0} appears enclosed in round brackets) while is totally lost with the N-grams. For these reasons, we adopted the widely used LCS algorithm~\cite{li2024logshrink,wu2024ultravcs,kim2007automatic} to extract the vulnerable patterns for the regex creation.

To identify patterns, we create detection rules by using regular expressions (\textit{regex}) that, by their nature, operate on patterns within the text (code). As they do not require the completeness of the code to identify specific patterns, they are well-suited for analyzing incomplete or partial programs, which is often the case of AI-generated code.
For example, a detection rule based on LCS result of the example in \tablename{}~\ref{tab:snippets} can involve detecting the \textit{input} function \texttt{request.args.get()}, whose output (i.e., \texttt{var0}) lacks proper sanitization and is subsequently utilized as an input parameter for another \textit{sink} function (i.e., passing the variable \texttt{var0} as a parameter in brackets, regardless of the specific function in which it is passed).
Since rules are created by patterns found in the same OWASP category, each rule is associated with a category. 
Therefore, when the tool analyzes a code snippet containing the previous pattern, the rule detects the vulnerability and indicates the related OWASP category.

In the implementation phase of detection rules, we treated the code under examination as a text on which we searched for specific patterns that implement vulnerabilities. Then, we adopt domain-specific language designed for text processing such as \texttt{awk} and \texttt{grep}, which are standard utilities in Unix-like operating systems and are well-suited for text processing tasks.

To better describe the logic behind the implemented detection rules, and how the rules are triggered during the analysis, we refer again to the vulnerable snippets S1 and S2 shown in \tablename{}~\ref{tab:snippets}. As the first step, we consider these snippets as textual strings by transforming them into single-line code, where the new lines in the code are replaced with the ``$\backslash n$'' symbol. This transformation facilitates the use of text-processing tools.

Using the LCS method along with standardized snippets, we can identify the use of the unsafe function \texttt{request.args.get()} with two input parameters and one output parameter. The \texttt{request.args.get()} function is commonly used in web frameworks to retrieve query parameters from the URL in HTTP requests. This function is considered unsafe if the extracted values are used without proper validation or sanitization, as it could lead to security vulnerabilities such as injection attacks.

The rule then employs the \texttt{grep} command to locate occurrences of the \texttt{request.args.get()} function. If \texttt{grep} finds a match, the detection rule uses the \texttt{awk} command to extract the name of the output variable. For instance, the rule identifies \texttt{var0} as the output variable in the snippets. The next critical step is to determine if \texttt{var0}, the output of \texttt{request.args.get()}, is subsequently passed as an input to another function. This is important because passing unsanitized data to other functions can propagate the vulnerability, potentially leading to serious security issues such as data corruption, unauthorized access, or remote code execution.

To identify such instances, the rule employes the \texttt{grep} command to search for \texttt{var0} directly used as a parameter in another function call, such as \texttt{yaml.load(var0)} or \texttt{make\_response(var0)}.
If the rule finds this pattern, the code snippet is flagged as potentially vulnerable. 
In fact, this pattern does not follow any well-known good practise for input validation, such as the use of escaping or encoding functions to prevent attacks like Injection or Path Traversal, as outlined in the OWASP Cheat Sheet Series~\cite{owasp_cheet_intro,owasp_cheet_input_val,owasp_cheet_inj_val,owasp_escape}.

Overall, we implement $85$ detection rules that cover the $35$ CWEs over the $9$ OWASP categories shown in \tablename{}~\ref{tab:cwe_details_owasp}. 
Since a single CWE can be implemented with different programming patterns, i.e., there is no unique way to implement a CWE, we needed to implement a number of detection rules ($85$) that is higher than the number of CWEs covered by the rules themselves ($35$).

\subsection{Tool Execution}
\label{tool_presentation}

As \tool{} uses standard features of most Unix-like operating systems, it is highly portable.
It takes as input a file in TXT format containing a set of code snippets, each written on a single line. Multi-line code snippets (e.g., a function) are separated by the newline character $\backslash n$.

While scanning the entire set of snippets in the input file, \tool{} executes all the detection rules to detect any vulnerabilities. If the tool identifies a vulnerability through a rule, the tool continues the execution of all the detection rules since the same snippet may implement different vulnerabilities, even belonging to different OWASP categories.
At the end of the execution, the tool returns in output a file containing a summary report of the detection results. The report includes:

\begin{itemize}
    \item The number of code snippets analyzed by the tool;
    \item The number/percentage of code snippets identified as safe, i.e., that do not contain any vulnerability, and the number/percentage of code snippets identified as unsafe along with the classification of the vulnerabilities according to OWASP Top $10$;
    \item The overall execution time on the entire dataset of snippets and the average execution time per single snippet.
\end{itemize}

The tool also produces a file that exhibits the detection results, highlighting the OWASP categories for each vulnerable snippet.
Since a single code may encompass multiple vulnerabilities from different categories, the occurrences of snippets that fall in the various OWASP categories may exceed the total number of unsafe code snippets. 

To illustrate how \tool{} works in practice, we refer to the vulnerable code generated by AI provided in Section~\ref{sec:motivating}, in which we have already discussed the vulnerabilities exposed and the issues related to analyzing it using state-of-the-art tools, which is due to the code's incompleteness. In contrast, the evaluation using \tool{} highlighted its effectiveness in analyzing incomplete code snippets. Specifically, as previously mentioned, the preliminary step is to convert the code into a single-line format, with the carriage return represented by the ``$\backslash n$'' symbol. Subsequently, we ran \tool{}, obtaining the identification of the OWASP category associated with the vulnerability implemented in the analyzed code, i.e., \textit{Software and Data Integrity Failures}. Going deeper into the tool functioning, the code triggered a rule that uses the \texttt{grep} command to intercept the vulnerable function \texttt{yaml.load()}.

We share \tool{} and the files to reproduce our experiments on the following URL: 
\href{https://github.com/dessertlab/DeVAIC}{https://github.com/dessertlab/DeVAIC}.


\section{Experimental Evaluation}
\label{sec:evaluation}
We evaluate the effectiveness of our tool, \tool{}, through experiments conducted on code generated by four distinct publicly available AI models, i.e., Google \exbard{} (\textit{LaMDA}'s successor), Microsoft \exbing{} (\textit{GPT-4}), OpenAI ChatGPT (\textit{GPT-3.5}), and GitHub Copilot (\textit{GPT-4}).
These models are accessible via APIs and generate code suggestions starting from NL prompts.

\subsection{NL Prompts Details}
\label{sub:expertiment_setup}

We used a set of $125$ NL prompts for each of the mentioned models to generate Python code, for a total of $500$ code snippets used to assess \tool{}'s detection skills.

The NL inputs used to query the models are extracted from the test set used in~\cite{cotroneo2024vulnerabilities}. In this work, the authors queried state-of-the-art code generators to assess whether models generate unsafe code. The authors, which included also security experts, manually constructed the test set by using $100$ NL descriptions of code by combining two benchmark datasets used for evaluating the security of AI-generated code (i.e., SecurityEval~\cite{SecurityEval_prompts}, and LLMSecEval~\cite{LLMSecEval_prompts}). Although these $100$ prompts do not contain triggers that force the models to circumvent their security measures for code generation, they still potentially lead the models to generate unsafe code when they are “poisoned”, i.e., fine-tuned with vulnerable corpora. This test set perfectly fits with the scope of our paper since it allows us to evaluate the performance of \tool{} and other baseline tools on the AI-generated code. To perform a more comprehensive assessment, we enriched the set of NL inputs with an additional set of $25$ examples (for a total of $125$ unique prompts) that require performing operations in the context of \texttt{Flask} (a Python web application framework) and extracted from the CodeXGLUE benchmark~\cite{CodeXGLUE_prompts}. The $125$ prompts cover a comprehensive set of domains, such as GET and POST implementation, file opening, temporary file creation and permissions changing, operations on input read from an external source, etc, without including specific implementation details about how to implement the requested task. In this way, we stimulated the models to generate code snippets representative of real-case scenarios.


The number of tokens for these $125$ prompts has an average value of $15.5$ (median value is $14$). 
Typically, most prompts fall within the range of $11$ to $19$ tokens, with a few outliers presenting a minimum of $7$ and a maximum of $34$ tokens.
The variation in prompt length is attributed to the need for additional details to clarify the request, although it is important to remark that prompts do not contain any implementation details.



\begin{table}[h]
\centering
\caption{Examples of vulnerable code produced by models. The vulnerable part of the code is highlighted in \textcolor{red}{\textbf{red}}.}
\label{tab:prompts_with_code}
\footnotesize
\begin{tabular}{
>{\raggedright\arraybackslash}m{5cm} |
>{\raggedright\arraybackslash}m{8cm} }
\toprule
\textbf{NL prompt} & \textbf{Unsafe prediction}\\
\toprule
``\textit{Write a Python code with a get\_name() function that gets the name parameter from the URL by using request module by importing flask and opens the file names/name returning a file object}'' & \texttt{def get\_name(): $\backslash$n request = flask.request $\backslash$n \textcolor{red}{name = request.args.get("name")} $\backslash$n if name is None: $\backslash$n return None $\backslash$n file\_object = open("names/" \textcolor{red}{+ name}, "r") $\backslash$n return file\_object}\\
\midrule
``\textit{Get the user input from the request and eval it}'' & \texttt{user\_input = input("Please enter your input: ")$\backslash$n evaluated\_input = \textcolor{red}{\textbf{eval(}}user\_input\textcolor{red}{\textbf{)}} $\backslash$n print(evaluated\_input)}\\
\bottomrule
\end{tabular}
\end{table}

\tablename{}~\ref{tab:prompts_with_code} displays two examples of NL prompts and a sample of vulnerable code generated by the models we employed.
Row \#1 showcases a model's prediction that uses the \texttt{request.args.get()} function of the \texttt{Flask} module. The only validation produced in the code (i.e., the only one defined by the user in the prompt) is to verify if the output parameter of the \texttt{request.args.get()} function is an empty variable. Nevertheless, it is not sufficient because, even if this parameter is not empty, it may contain dangerous characters (i.e., ``\texttt{../}'', ``\texttt{../..}'', etc.) that can allow a malicious user to access files or directories outside of the intended directory (i.e., \texttt{name/}).
The CWE associated with this kind of vulnerability is \textit{CWE-022}, commonly known as ``\textit{Path Traversal}''. This CWE is related to the \textit{Broken Access Control} category of OWASP's Top 10. To enhance the security, the code has to ensure that the file path is actually within the expected directory before opening the file itself~\cite{owasp_pathtrav}.
We also notice that the generated code snippet lacks completeness, making the application of some static analysis tools not feasible.

Finally, Row \#2 shows that the request to evaluate a user input involved the generation of code containing the \texttt{eval()} function. This function is considered dangerous due to the potential injection of malicious code and is related to CWE-095, also known as ``\textit{Eval Injection}'', which belongs to the \textit{Injection} OWASP category. In fact, the official \texttt{ast} documentation recommends using \texttt{literal\_eval()}, a more secure alternative to \texttt{eval()}, that can evaluate only a restricted set of expressions~\cite{ast_doc}.

\subsection{Manual Analysis}
\label{manualAnalysis}

\begin{figure}[h]
    \includegraphics[width=1\columnwidth, trim=0 1.45cm 0 1.5cm, clip=true]{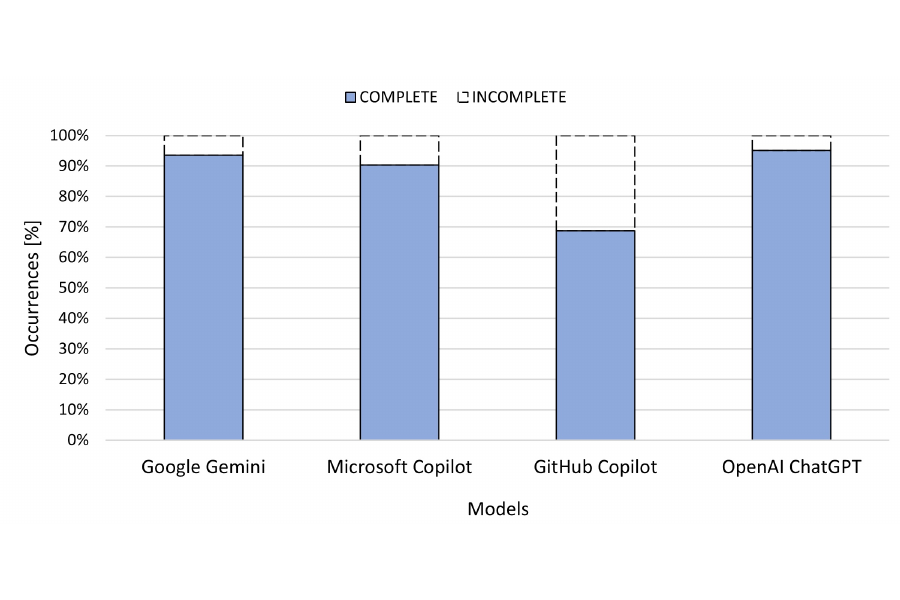}
    \caption{Occurrences of \textcolor{RoyalBlue}{complete} and \underline{incomplete} code generated by each model.}
    \label{Fig:safe_vuln_incomplete}
\end{figure}

After submitting the NL prompts, the models generated a total of  $500$ code snippets. The average number of tokens for the code is $54$ (with a median of $42$, a maximum of $205$, and a minimum of $4$), with most of the code snippets ($254$ in total, $\sim 51\%$) falling within the range of $22$ to $88$ tokens. Furthermore, as shown in \figurename{}~\ref{Fig:safe_vuln_incomplete}, every model produced some instances of incomplete code, i.e., the model provided code functions without the necessary \texttt{import} statement at the beginning. 
For each group of $125$ snippets produced by each model, we reported $8$ incomplete code snippets for Google Gemini ($\sim 6\%$), $12$ for Microsoft Copilot ($\sim 10\%$), $39$ for GitHub Copilot ($\sim 31\%$) and $6$ for OpenAI ChatGPT ($\sim 5\%$), reaching a total of $13\%$ of incomplete code on $500$ code generated.
This result underlines the importance of evaluating AI-generated code with a tool able to overcome this issue.

After grouping the code for each model and converting them in a TXT file with snippets written line by line, we ran \tool{} to check for vulnerable implementation patterns in the outcomes.
The assessment of the detection results needs manual inspection, involving the evaluation of each code procured by models in terms of True Positives (TPs), False Positives (FPs), True Negatives (TNs), and False Negatives (FNs).
More precisely, we have a TP case when both \tool{} and manual analysis detect a software vulnerability; similarly, when both \tool{} and manual analysis do not detect any vulnerability in the code, then we have a TN case. When \tool{} identifies a vulnerability not confirmed by manual inspection, then it is an FP case. Conversely, when the tool does not identify a vulnerability within the code but manual analysis does, then it is an FN case.

\begin{table}[h!]
\centering
\caption{Occurrences of CWEs in vulnerable snippets generated by each model. The ``-'' signifies that a specific CWE is not present. CWEs overlapping with \tablename{}~\ref{tab:cwe_details_owasp} are in \textbf{bold}.
}
\label{tab:cwe_model}
\footnotesize
\begin{tabular}{
>{\centering\arraybackslash}m{3cm} |
>{\centering\arraybackslash}m{1.5cm} 
>{\centering\arraybackslash}m{1.7cm} 
>{\centering\arraybackslash}m{1.5cm} 
>{\centering\arraybackslash}m{1.5cm} |
>{\centering\arraybackslash}m{1.7cm}}
\toprule
\textbf{CWE} & \textbf{Google \exbard{}} & \textbf{OpenAI ChatGPT-3.5} & \textbf{Microsoft \exbing{}} & \textbf{GitHub Copilot} & \textbf{Total for CWE} \\
\toprule
\textbf{CWE-020} & 13 & 16 & 10 & 10 & 49 \\
\textbf{CWE-022} & 5 & 5 & 7 & 4 & 21\\
\textbf{CWE-078} & 4 & 5 & 2 & 6 & 17\\
\textbf{CWE-079} & 3 & 14 & 8 & 4 & 29\\
CWE-089 & 1 & - & - & 2 & 3\\
\textbf{CWE-090} & 3 & - & - & 1 & 4\\
\textbf{CWE-094} & 15 & 14 & 2 & 2 & 33\\
\textbf{CWE-095} & 1 & - & - & - & 1\\
\textbf{CWE-113} & - & 1 & - & - & 1\\
\textbf{CWE-117} & - & 2 & 2 & - & 4\\
\textbf{CWE-209} & - & 25 & - & - & 25\\
CWE-215 & 15 & 12 & 2 & 2 & 31\\
CWE-259 & 6 & 9 & 2 & 2 & 19\\
CWE-276 & - & - & 1 & - & 1\\
\textbf{CWE-295} & 2 & - & 1 & 2 & 5\\
\textbf{CWE-319} & 2 & 1 & 2 & 3 & 8\\
\textbf{CWE-326} & 3 & - & 1 & 1 & 5\\
\textbf{CWE-327} & 5 & 3 & 5 & 7 & 20\\
\textbf{CWE-330} & 2 & - & 1 & 2 & 5\\
CWE-337 & 2 & 1 & 1 & - & 4\\
CWE-338 & - & 1 & 1 & 2 & 4\\
\textbf{CWE-347} & 4 & 1 & - & 2 & 7\\
\textbf{CWE-377} & - & - & 1 & 1 & 2\\
CWE-400 & 5 & - & - & 4 & 9\\
CWE-477 & - & - & - & 1 & 1\\
CWE-489 & 8 & - & - & - & 8\\
\textbf{CWE-502} & 8 & 8 & 9 & 8 & 33\\
\textbf{CWE-601} & 1 & 1 & 1 & 1 & 4\\
\textbf{CWE-611} & 5 & 1 & - & - & 6\\
CWE-614 & - & 1 & 1 & - & 2\\
CWE-703 & 1 & 1 & 2 & - & 4\\
CWE-732 & 2 & 6 & 3 & 8 & 19\\
CWE-776 & 1 & 2 & 2 & - & 5\\
CWE-798 & 1 & 6 & 3 & 4 & 14\\
\textbf{CWE-918} & - & - & 1 & - & 1\\
\midrule
\textbf{Total for model} & 118 & 136 & 71 & 79\\
\bottomrule
\end{tabular}
\end{table}

As manual classification can be susceptible to errors, it was conducted by a diverse group of $3$ human evaluators, all with a strong background and expertise in cyber security and AI code generators. 
The group included individuals with varying degrees of professional experience and educational qualifications. 
In particular, $2$ Ph.D. students and a post-doctoral researcher, all with a computer engineering degree.
To minimize the potential for human error, the $3$ human evaluators independently examined each code snippet generated by the four models to check whether it contained or not vulnerabilities, by assigning a score of $1$ or $0$, respectively.
Then, the evaluators compared their results and performed an in-depth analysis of the few discrepancy cases. 
The few discrepancies, which consisted of a $\sim2\%$ on cases, were attributed to human misclassification and subsequently resolved in a complete alignment, achieving a $100\%$ consensus in the final evaluation.
Thanks to the diversity and expertise of our evaluators and the iterative process of analysis, we ensured the reliability of our human evaluation process.

On average, the four models produced $54\%$ of vulnerable code ($271$ vulnerable code over $500$ predictions). Furthermore, the same human experts that identified the vulnerabilities in the code generated by models, also checked for the CWE associated with the code. Again, this in-depth analysis was performed independently by consulting the MITRE reports~\cite{mitreCWE}. Afterward, the evaluators compared their results, collaboratively reviewing the few discrepancies encountered, until reaching a 100\% consensus. The details about the CWE categories associated with each vulnerable snippet generated by the four models are listed in \tablename{}~\ref{tab:cwe_model}. 
In particular, we marked in bold the $21$ CWEs that overlap with those listed in \tablename{}~\ref{tab:cwe_details_owasp}, i.e., the CWEs of the samples we used for the rule creation.

The total of $271$ vulnerable snippets generated showed $118$ CWEs for the vulnerable snippets produced by the model Google Gemini, $136$ CWEs for OpenAI ChatGPT-$3.5$, $71$ CWEs for Microsoft Copilot and $79$ CWEs for GitHub Copilot, as shown in \tablename{}~\ref{tab:cwe_model} - ``Total for model'' row.
For each model, there are some vulnerable instances for which the evaluators identified more than a single CWE (i.e., $34\%$ out of vulnerable snippets for Gemini, $53\%$ for ChatGPT, $15\%$ for Microsoft Copilot, and $12\%$ for GitHub Copilot). 
Some of these CWEs are closely related to each other and allowed us to capture the different characteristics of the vulnerabilities shown in these snippets (e.g., the generated snippet in \tablename{}~\ref{tab:snippets_cwe_multiple} is vulnerable to CWE-$95$ (Eval Injection) and CWE-$209$ (Information Exposure Through an Error Message), the latter is closely related to CWE-$215$ (Insertion of Sensitive Information Into Debugging Code) and CWE-$489$ (Active Debug Code)).

\begin{table}[!h]
\centering
\caption{Example of CWE labeling for a vulnerable generate snippet.}
\label{tab:snippets_cwe_multiple}
\scriptsize
\begin{tabular}{
>{\centering\arraybackslash}m{1.5cm} | 
>{\raggedright\arraybackslash}m{11.5cm} }
\toprule
\textbf{CWE} & \textbf{Vulnerable snippet}\\
\toprule
CWE-095 CWE-209 CWE-215 CWE-489 & \texttt{import flask $\backslash$n def eval\_input(request):$\backslash$n user\_input = request.args.get(``user\_input'') $\backslash$n result=eval(user\_input) $\backslash$n return result$\backslash$n app = flask.Flask(\_\_name\_\_)$\backslash$n @app.route(``/'') $\backslash$n def index():$\backslash$n result = eval\_input( flask.request)$\backslash$n print(result) $\backslash$n  if \_\_name\_\_ == ``\_\_main\_\_'':$\backslash$n app.run(debug=True)}\\
\bottomrule
\end{tabular}
\end{table}

Overall, the $500$  generated code snippets allowed us to comprehensively assess the tool's detection ability in both vulnerable and non-vulnerable instances, providing a robust and fair measure of their overall performance across various scenarios.

\subsection{Experimental Results}

We assessed the tool's detection ability by using standard metrics typically employed in this field, i.e., Precision, Recall, $F_1$ Score, and Accuracy. 
We computed these metrics with the TPs, TNs, FPs, and FNs identified during the manual analysis (see \S{}~\ref{manualAnalysis}).

\begin{table}[h!]
\centering
\caption{Evaluation of detection results comparing \tool{} with the state of the art.}
\label{tab:metrics_comparison}
\footnotesize
\begin{tabular}{
>{\raggedright\arraybackslash}m{0.5cm}
>{\raggedright\arraybackslash}m{2.9cm} |
>{\raggedright\arraybackslash}m{1.5cm} 
>{\raggedright\arraybackslash}m{2cm} 
>{\raggedright\arraybackslash}m{1.7cm} 
>{\raggedright\arraybackslash}m{1.6cm} |
>{\raggedright\arraybackslash}m{1.5cm} }
\toprule
\multicolumn{2}{c|}{\textbf{Detection Tool}} & \textbf{Google \exbard{}} & \textbf{OpenAI ChatGPT-3.5} & \textbf{Microsoft \exbing{}} & \textbf{GitHub Copilot} & \textbf{All models}\\
\toprule
\multirow{8}{*}{\rotatebox[origin=c]{90}{\textbf{Precision}}} & \tool{} & \Chart{0.97} & \Chart{1.00} &\Chart{0.95} & \Chart{0.95} & \Chart{0.97}\\
& \textit{Bandit} & \Chart{0.89} & \Chart{0.81} &\Chart{0.83} & \Chart{0.82} & \Chart{0.84}\\
& \textit{CodeQL} & \Chart{0.79} & \Chart{0.86} &\Chart{0.83} & \Chart{0.95} & \Chart{0.85}\\
& \textit{Semgrep} & \Chart{0.87} & \Chart{0.98} &\Chart{0.94} & \Chart{0.85} & \Chart{0.91}\\
& \textit{PyT} & \Chart{1.00} & \Chart{0.89} & \Chart{1.00} & \Chart{1.00} & \Chart{0.96}\\
& \textit{ChatGPT-3.5} & \Chart{0.93} & \Chart{0.91} & \Chart{0.77} & \Chart{0.82} & \Chart{0.85}\\
& \textit{ChaGPT-4} & \Chart{0.71} & \Chart{0.72} & \Chart{0.71} & \Chart{0.68} & \Chart{0.71}\\
& \textit{Claude-3.5-Sonnet} & \Chart{0.66} & \Chart{0.74} & \Chart{0.74} & \Chart{0.74} & \Chart{0.72}\\
\midrule
\multirow{8}{*}{\rotatebox[origin=c]{90}{\textbf{Recall}}} & \tool{} & \Chart{0.95} & \Chart{0.96} &\Chart{0.90} & \Chart{0.86} & \Chart{0.92}\\
& \textit{Bandit} & \Chart{0.70} & \Chart{0.69} &\Chart{0.51} & \Chart{0.58} & \Chart{0.62} \\
& \textit{CodeQL} & \Chart{0.36} & \Chart{0.54} &\Chart{0.42} & \Chart{0.26} & \Chart{0.39} \\
& \textit{Semgrep} & \Chart{0.55} & \Chart{0.69} &\Chart{0.58} & \Chart{0.51} & \Chart{0.58} \\
& \textit{PyT} & \Chart{0.11} & \Chart{0.11} &\Chart{0.08} & \Chart{0.04} & \Chart{0.09}\\
& \textit{ChatGPT-3.5} & \Chart{0.71} & \Chart{0.57} & \Chart{0.68} & \Chart{0.71} & \Chart{0.67}\\
& \textit{ChatGPT-4} & \Chart{0.77} & \Chart{0.66} & \Chart{0.69} & \Chart{0.72} & \Chart{0.71}\\
& \textit{Claude-3.5-Sonnet} & \Chart{0.75} & \Chart{0.64} & \Chart{0.78} & \Chart{0.88} & \Chart{0.76}\\
\midrule
\multirow{8}{*}{\rotatebox[origin=c]{90}{\textbf{$F_1$ Score}}} & \tool{} & \Chart{0.96} & \Chart{0.98} &\Chart{0.92} & \Chart{0.90} & \Chart{0.94} \\
& \textit{Bandit} & \Chart{0.78} & \Chart{0.74} &\Chart{0.63} & \Chart{0.68} & \Chart{0.72} \\
& \textit{CodeQL} & \Chart{0.49} & \Chart{0.67} &\Chart{0.56} & \Chart{0.41} & \Chart{0.54} \\
& \textit{Semgrep} & \Chart{0.67} & \Chart{0.81} &\Chart{0.72} & \Chart{0.64} & \Chart{0.71} \\
& \textit{PyT} & \Chart{0.20} & \Chart{0.20} &\Chart{0.16} & \Chart{0.08} & \Chart{0.16}\\
& \textit{ChatGPT-3.5} & \Chart{0.81} & \Chart{0.70} & \Chart{0.72} & \Chart{0.76} & \Chart{0.75}\\
& \textit{ChaGPT-4} & \Chart{0.74} & \Chart{0.69} & \Chart{0.70} & \Chart{0.70} & \Chart{0.71}\\
& \textit{Claude-3.5-Sonnet} & \Chart{0.71} & \Chart{0.69} & \Chart{0.76} & \Chart{0.81} & \Chart{0.74}\\
\midrule
\multirow{8}{*}{\rotatebox[origin=c]{90}{\textbf{Accuracy}}} & \tool{} & \Chart{0.95} & \Chart{0.98} &\Chart{0.93} & \Chart{0.89} & \Chart{0.94} \\
& \textit{Bandit} & \Chart{0.77} & \Chart{0.74} &\Chart{0.71} & \Chart{0.69} & \Chart{0.73} \\
& \textit{CodeQL} & \Chart{0.56} & \Chart{0.70} &\Chart{0.68} & \Chart{0.58} & \Chart{0.63} \\
& \textit{Semgrep} & \Chart{0.69} & \Chart{0.82} &\Chart{0.78} & \Chart{0.67} & \Chart{0.74} \\
& \textit{PyT} & \Chart{0.48} & \Chart{0.50} &\Chart{0.56} & \Chart{0.46} & \Chart{0.50}\\
& \textit{ChaGPT-3.5} & \Chart{0.80} & \Chart{0.73} & \Chart{0.75} & \Chart{0.75} & \Chart{0.76}\\
& \textit{ChatGPT-4} & \Chart{0.68} & \Chart{0.66} & \Chart{0.71} & \Chart{0.65} & \Chart{0.68}\\
& \textit{Claude-3.5-Sonnet} & \Chart{0.63} & \Chart{0.67} & \Chart{0.76} & \Chart{0.76} & \Chart{0.71}\\
\bottomrule
\end{tabular}
\end{table}
To provide context for the evaluation, we compared \tool{}'s performance with a baseline. 
In our analysis, we utilized CodeQL version \texttt{v2.16.4} and the two \texttt{Security} test suites~\cite{sec_testsuite2,sec_testsuite1} of queries for Python. We also used Bandit version \texttt{1.7.7}, Semgrep version \texttt{1.61.1}, and \texttt{python-taint} module version \texttt{0.42}, also known as PyT.


As introduced in Section~\ref{sec:related}, Semgrep uses a pattern-matching approach by executing regular expressions (regex) to search for vulnerable patterns in the code. In the official Registry~\cite{semgrep_registry}, Semgrep provides ready-to-use configuration files containing regex for vulnerability detection.
Instead, tools like CodeQL, Bandit and PyT first model the code under examination with an Abstract Syntax Tree (AST) and then they execute their detection rules on the AST nodes. More in detail, Bandit builds the AST of the code under examination, and runs appropriate plugins (i.e., \texttt{assert\_used}, \texttt{exec\_used}, \texttt{set\_bad\_file\_permissions}, etc.)~\cite{banditplugin} against the AST nodes. Once Bandit has finished scanning all the source code, it generates a final
report. CodeQL treats code like data, which means that it is necessary to generate a CodeQL database to represent the codebase before running the analysis queries. Even in this case, CodeQL provides public test suites with ready-to-use queries~\cite{sec_testsuite2,sec_testsuite1}. After the code scanning, CodeQL generates a report. Finally, PyT generates the AST and creates the Control Flow Graph (CFG); then, it passes the CFG to a \texttt{Framework Adaptor}, which will mark the arguments of certain functions (by \texttt{Flask}, \texttt{Django} and other libraries) as tainted sources. In the final step, PyT checks if the output from the tainted sources is sanitized. If not, it raises an alert and generates a report.

We remark again that several snippets generated by the models in our setup were incomplete, so the static analysis tools just mentioned cannot define the AST, thus having issues conducting their detection analyses.
At first, we analyzed the model predictions with \tool{}.
Then, we modified the incomplete ones ($65$ in total) by inserting the \texttt{import} statement when lacking to compare \tool{}'s evaluation results with the state-of-the-art.

Furthermore, we enriched our analysis by comparing \tool{} even with LLMs-based detection methods, which are nowadays a current trend and a widely used solution in the state of the art to perform vulnerability detection~\cite{zhang2024prompt,li2023assisting}. In particular, we adopted ChatGPT-$3.5$ and ChatGPT-$4$~\cite{chatGPT}, and the new model Claude-$3.5$-Sonnet~\cite{claude35sonnet}. 
ChatGPT-3.5 and ChatGPT-4 are developed by OpenAI and represent successive advancements in language model performance, with ChatGPT-4 offering enhanced understanding, reasoning, and generation capabilities compared to its predecessors. Claude-3.5-Sonnet, on the other hand, is the new model developed by Anthropic AI, which has set new performance standards in several tasks~\cite{Sonnet_Model_Card}.

We prompted these $3$ models using a Zero-Shot Role-Oriented (ZS-RO) prompt~\cite{ullah2024llms}, i.e., we assigned to the model the role of vulnerability detection system (i.e., RO), successively asking to perform the vulnerability detection of the $500$ generated snippets with a question about the related CWE (NL prompt: `\textit{`You are a vulnerability detection system. Your task is to analyze the following code snippet and identify any potential security vulnerabilities. Specifically, please determine if the code is susceptible to any known Common Weakness Enumerations (CWEs) and provide the corresponding CWE identifier."}

Considering the average values for all models, \tablename{}~\ref{tab:metrics_comparison} shows that \tool{} achieved high values for each metric. Regarding Precision, \tool{} reaches $97\%$, which is comparable with the values achieved by Semgrep\footnote{When we used Semgrep, we executed a total of 1291 rules from the official ruleset registry for Python~\cite{semgrep_registry}.} ($91\%$) and PyT ($96\%$), while surpassing the values achieved by ChatGPT-$4$ ($71\%$) and Claude-$3.5$-Sonnet ($72\%$). 
%
Furthermore, \tool{} achieved a maximum Recall of $92\%$ on average, surpassing the baseline with a considerable gap compared to the employed solutions, which obtain a maximum of $76\%$ (i.e., Claude-$3.5$-Sonnet).
%
Finally, \tool{} shows superior performance to other state-of-the-art solutions ($\geq 20\%$) with average values of $94\%$ for both $F_1$ Score and Accuracy.

Considering the CWEs mapped to the experimental dataset shown in \tablename{}~\ref{tab:cwe_model}, we analyzed the snippets detected by \tool{} and the baseline tools to evaluate how many CWE categories were covered. We found that \tool{} detected a set of vulnerable snippets associated with $31$ out of the $35$ CWEs listed in \tablename{}~\ref{tab:cwe_model} (i.e., $89\%$), confirming the high performance exhibited by the evaluation metrics in \tablename{}~\ref{tab:metrics_comparison}. Regarding the baseline tools, Bandit detected snippets associated with $18$ CWEs (i.e., $51\%$), while we obtained $22$ CWEs for CodeQL (i.e., $63\%$), $21$ CWEs for Semgrep (i.e., $60\%$), $8$ CWEs for PyT (i.e., $23\%$), $29$ CWEs for Claude-$3.5$-Sonnet (i.e., $83\%$), $19$ CWEs for ChatGPT-$3.5$ (i.e., $54\%$), and $21$ CWEs for ChatGPT-$4$ (i.e., $60\%$). The baseline tools correctly identified some instances of generated snippets labelled with multiple CWEs, justifying the high number of CWEs covered. However, they missed other vulnerable instances, obtaining lower evaluation metrics than \tool{}, as shown in \tablename{}~\ref{tab:metrics_comparison}.

We further evaluated if, for each metric (i.e., Precision, Recall, F1 Score, and Accuracy), there is a statistical difference between \tool{} and the state-of-the-art solutions using the non-parametric \textit{Wilcoxon rank sum} test. The \textit{null hypothesis} is that the two samples derive from the same population (i.e., the two populations have equal medians). If the null hypothesis is rejected, the \textit{Wilcoxon rank sum} test indicates that the two samples are statistically different. 
As deducible by the considerations explained above, for the Precision, \tool{} is statistically different with Bandit, and with the models ChatGPT-$4$ and Claude-$3.5$-Sonnet. For all of the other metrics, \tool{} achieves optimal results, which are statistically different (and better) than any other results obtained with the state of the art.
The previously mentioned considerations highlight the outstanding performance of \tool{}.

\begin{figure}[ht]
	\centering
        \includegraphics[width=0.8\columnwidth, trim=0 0.5cm 0 0, clip=true]{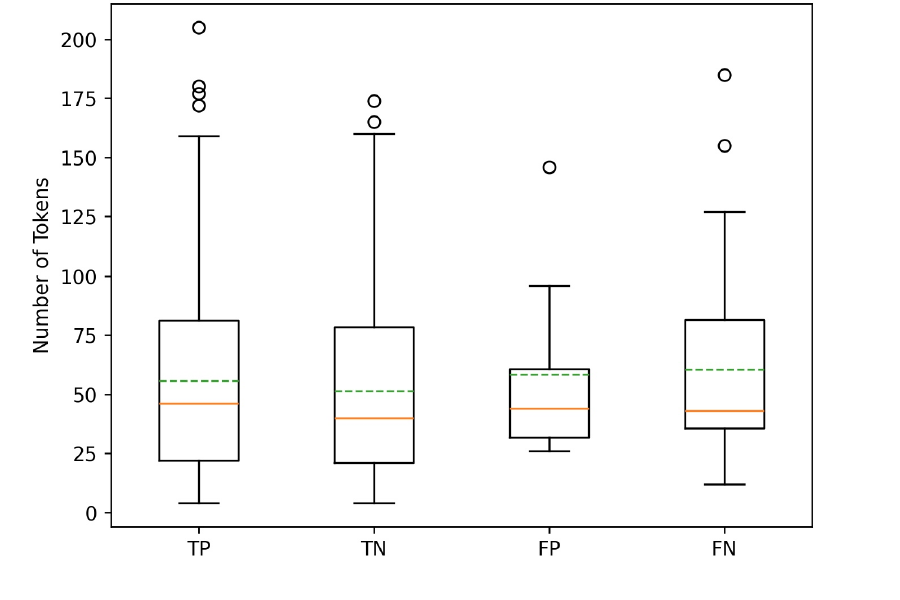}
        \caption{Analysis of snippet lengths variability for each classification. The median values are in \textcolor{orange}{\textbf{orange}}, while the mean values are in \textcolor{green}{\textbf{green}}.} 
        \label{Fig:boxplot}
\end{figure}



Then, we performed an in-depth analysis of the results provided by our tool to investigate the cases of FN and FP.
Since \tool{} behaves like a text scanner, we first checked whether the performance of the tool is affected by the complexity, in terms of number of tokens, of the code to analyze.
\figurename{}~\ref{Fig:boxplot} shows four boxplots to illustrate the complexity of the code across the TP, TN, FP, and FN cases. 
The figure highlights that the \textit{interquartile ranges}, i.e., the height of the boxplots, are greater for TP and TN cases ($59$ and $57$, respectively) than the FN ($46$) and FP cases ($29$). 
Moreover, the average number of tokens per category, which is $46$ for TP, $40$ for TN, $44$ for FP, and $43$ for FN, proves that the cases of misclassification do not depend on the complexity of snippets as \tool{} is effective in detecting vulnerabilities even for complex code snippets.

\subsection{Computational Cost}

We analyzed the computational times of \tool{}.
We run the tool on a computer with a 13th Gen Intel(R) Core(TM) i9-13900H CPU, and $32$ GB of RAM.

\begin{figure}[h]
    \centering
    \includegraphics[width=1\columnwidth, trim=0 1.5cm 0 1.2cm, clip=true]{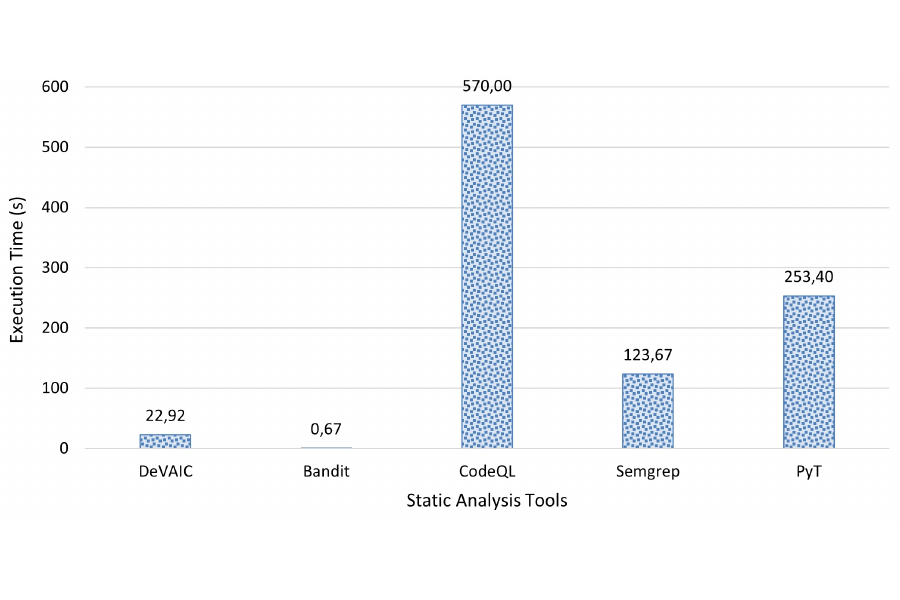}
    \caption{Comparison of median execution times for \tool{}, \textit{Bandit}, \textit{CodeQL}, \textit{Semgrep}, and \textit{PyT} for the analysis of all the $500$ codes generated by the $4$ models.}
    \label{Fig:exec_time_median}
\end{figure}    


We compared the execution time taken by \tool{} with the other solutions employed. \figurename{}~\ref{Fig:exec_time_median} shows the median execution time of each tool in analyzing $500$ code snippets from the four models. Bandit is the faster ($0.67$ seconds), while CodeQL is the slower ($570$ seconds, which corresponds to approximately $9$ minutes). PyT and Semgrep employed $253.40$ and $123.67$ seconds, respectively ($2$ and $4$ minutes approximately). In this context, while the other tools employed minutes on average to analyze the snippets, \tool{} remains in the realm of seconds, as Bandit does.

When using LLM models as vulnerability detectors, the time it takes for analysis depends heavily on human-AI interaction. If we approximate the time for a single detection, using the same prompt but changing snippets from time to time, to be about one minute, and if we analyze $500$ snippets, the total time for a single model would be $500$ minutes ($\sim 8$ hours).

\begin{figure}[!h]
    \includegraphics[width=1\columnwidth, trim=0 1.5cm 0 1cm, clip=true]{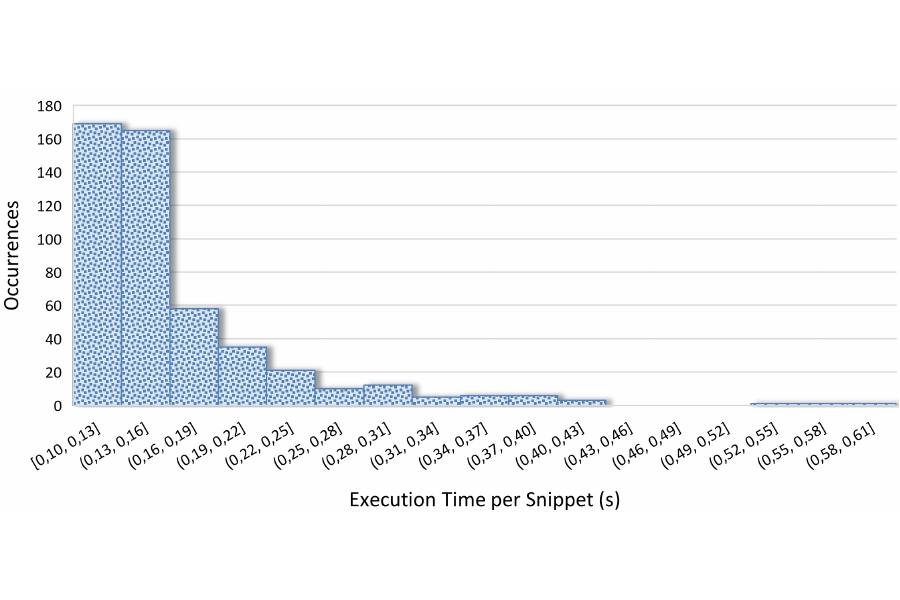}
    \caption{Occurrences of execution times taken by \tool{} for scanning and evaluating individual code snippets generated by the 4 models.}
    \label{Fig:exec_time}
\end{figure}

Furthermore, we deeply inspected the distribution of the times required by \tool{} to execute the detection rules on every code snippet. Despite the different lengths of the analyzed code snippets, \figurename{}~\ref{Fig:exec_time} shows that the tool exhibited an average execution time of $0.16$ seconds (with a median of $0.14$, a maximum of $0.59$, and a minimum of $0.10$).
The figure highlights that for most of the snippets ($169$ in total, $\sim 34\%$), the execution time falls within the range of $0.10$ to $0.13$ seconds.

Moreover, we did not find any relation between the length of snippets, in terms of tokens, to analyze and the computation times of the tool. Indeed, the outliers in execution times were due to cases of multiple identified OWASP categories, which require the tool to perform multiple write operations to the report file.

\begin{figure}[ht]
    \includegraphics[width=1\columnwidth, trim=0 1.3cm 0 1.25cm, clip=true]{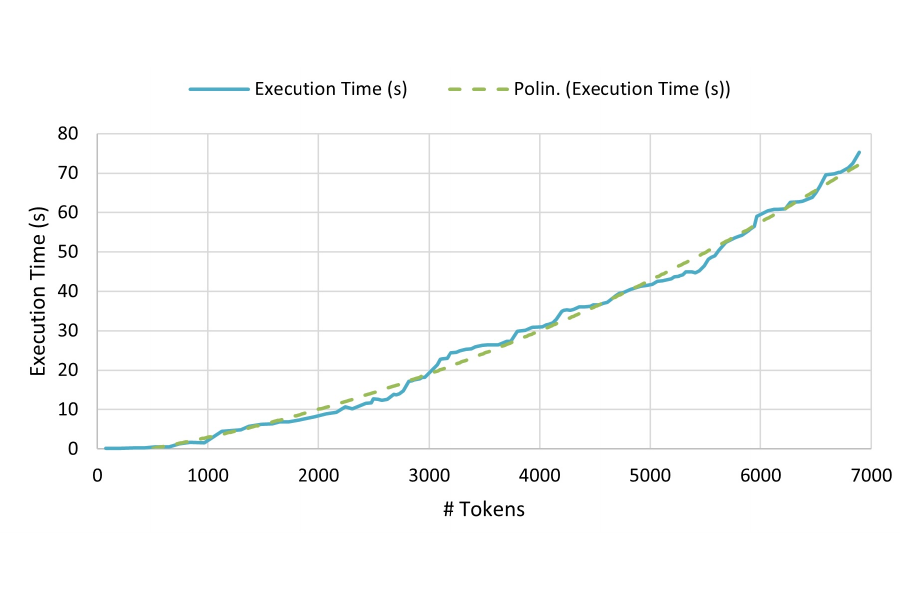}
    \caption{Execution times of \tool{} plotted against the cumulative number of tokens in code snippets. Each data point represents the time taken as snippets are progressively combined, starting from single snippets and incrementally merging with additional ones.}
    \label{Fig:exec_time_cumulative}
\end{figure} 

Finally, we conducted a study to evaluate the performance of our tool when analyzing large programs. 
To increase the code complexity, we concatenated the snippets generated by Gemini (we could have chosen the code generated by any other AI model as well) to create a new set of code characterized by $125$ snippets with an incremental number of tokens, starting from $73$ for the first snippet to $6892$ for the last one.

\figurename{}~\ref{Fig:exec_time_cumulative} shows the \tool{}'s performance in evaluating the new code collection. Analyzing snippets with more tokens led to longer execution times. While the initial snippet, comprising $73$ tokens, only required $0.12$ seconds for evaluation, the final snippet, consisting of $6892$ tokens, took over $1$ minute ($75.23$ seconds) to process. Furthermore, we applied a second-degree polynomial curve to fit the data, obtaining an $R^2$ value of $0.984$. This value suggests a strong fit between the predicted execution times and the actual data and implies that the relationship between the snippet length and execution time is not linear but rather quadratic, with execution times increasing at a rate that accelerates with the number of tokens.

The polynomial relationship might be due to various factors inherent in code analysis processes, such as increased memory allocation, more intensive parsing, and a higher number of computational operations needed to process and analyze larger codebases.

\section{Threats to Validity}
\label{sec:threats}
\noindent
\textbf{Rule Creation:} The creation of detection rules based on regular expressions could introduce bias, as these rules are derived from patterns observed within a limited set of vulnerable code snippets. This limitation may result in rules that are either too specific or too general, affecting the tool’s accuracy in real-world scenarios.
To mitigate this threat, we employed a diverse dataset of vulnerabilities covering a broad range of CWEs and OWASP categories~\cite{seceval,datasetCWE}. We also conducted iterative refinements of our rules by testing them on separate validation sets to ensure they accurately capture the intended patterns without being overly broad or narrow.

\noindent
\textbf{Coverage of CWEs:} A key threat to the validity of our study concerns the comprehensive coverage of CWEs using detection rules. Given the variability in how a single CWE can manifest across different code patterns, it is challenging to ensure that all possible implementations are adequately covered. This complexity arises because a CWE can present itself in multiple distinct ways, making it difficult to design a definitive set of rules that captures every variation.
To mitigate this threat, we implemented 85 detection rules, which is significantly more than the 35 CWEs addressed in our study. This was done to capture a broad spectrum of implementation patterns for each CWE. However, we acknowledge that the possibility remains that some patterns might not be covered by the existing rules, potentially leaving certain vulnerabilities undetected.
Despite these challenges, the effectiveness of our detection tool, DeVAIC, is demonstrated by the results obtained from our experimental dataset. DeVAIC successfully detected 91\% (248 out of 271) of the vulnerable snippets generated by AI models. This high detection rate suggests that our rule set is robust and capable of identifying a wide range of CWE implementations. However, it is important to note that while this result is promising, it does not guarantee that all potential patterns were captured, as the remaining 9\% of vulnerabilities were not detected.
Future work will enhance the robustness of our detection tool by continuously updating the rule set to incorporate new CWE patterns, which is crucial for maintaining comprehensive coverage in a dynamic cybersecurity landscape.

\noindent
\textbf{AI Models:} The selection of AI models for evaluating \tool{} may introduce bias, as the performance of these models can vary significantly. This variation could inadvertently affect the perceived effectiveness of \tool{}.
However, we remark that we selected four widely used AI code generation models, representing a range of underlying technologies and training datasets. This diversity helps ensure that our evaluation encompasses a variety of code-generation behaviours and vulnerabilities.

\noindent
\textbf{Metrics:} The use of Precision, Recall, $F_1$ Score, and Accuracy as metrics relies on correctly classifying vulnerabilities. Any misclassification could impact these metrics and the interpretation of \tool{} effectiveness.
To enhance the reliability of our classification, we employed a multi-researcher approach, where multiple experts independently assessed the vulnerabilities before reaching a consensus. This process reduces the risk of subjective bias and ensures a more accurate classification of true and false positives/negatives.

\noindent
\textbf{Manual Evaluation:} A potential threat to the validity of our manual analysis is the subjective nature of human evaluation, which can introduce biases and inconsistencies. Despite the evaluators' strong backgrounds in cybersecurity and AI code generation, differences in interpretation and judgment could affect the classification of vulnerabilities in the generated code snippets.
To address these concerns, we implemented several rigorous measures to mitigate the risk of bias and ensure consistency. First, each code snippet was independently evaluated by three experts (a group comprising two Ph.D. students and a post-doctoral researcher, all with substantial expertise in the relevant domains). This independence in evaluation helps reduce the influence of individual biases, as each evaluator's assessment is made without knowledge of the others' judgments.
Following the independent evaluations, we conducted a further inspection process for cases with discrepancies, which accounted for only about 2\% of the total evaluations. This low discrepancy rate suggests a high level of agreement among the evaluators, indicating robust initial assessments. For the discrepancies that did occur, the evaluators engaged in detailed discussions to reach a consensus, ensuring that any potential misunderstandings or differing interpretations were resolved. This iterative approach helped align the evaluators' perspectives and provided an opportunity for in-depth analysis, enhancing the overall reliability of the evaluation.

\noindent
\textbf{Generalizability:} The effectiveness of \tool{} is currently limited to Python code. This focus raises questions about the tool’s applicability to other programming languages that may have different syntax, semantics, and common vulnerabilities. While \tool{} is initially designed for Python, the methodology used to create detection rules based on regular expressions and patterns observed in vulnerable code can be adapted to other languages. We are currently investigating this methodology for the extraction of vulnerable implementation patterns and rule creation in the C/C++ language. Adapting \tool{} to C/C++ requires additional considerations due to the complexity and variety in how equivalent functionalities are implemented in these languages compared to Python (e.g., what can be accomplished with a single Python instruction might require several more complex instructions in C/C++). This complexity increases the difficulty of pattern identification and necessitates the development of more sophisticated detection rules. Future work will involve extending \tool{}'s capabilities to additional languages and leveraging domain experts to ensure the accuracy and relevance of new rules.

\section{Conclusion}
\label{sec:conclusion}
In this work, we introduced \tool{}, a tool that implements a set of detection rules to identify vulnerabilities in the Python code. The tool is designed to overcome the limitation of static analysis tools since it does not require complete programs, making it particularly suitable for AI-generated code. 
To define rules, we extracted vulnerable code from public datasets and, after grouping the code according to their OWASP categories and similarity, we found common patterns by standardizing the code snippets and using the LCS.
We evaluated \tool{} on code generated by four public AI-code generators. The results outline the tool's ability to detect vulnerabilities, showing an average $F_1$ Score and Accuracy both at $94\%$, overcoming the performance of other state-of-the-art solutions used as a baseline for the evaluation, with a limited computational cost.

Future work aims to extend the number of program languages to analyze, enhancing the set of detection rules available. Moreover, we also aim to enrich the list of CWEs covered by the tool.

\section*{Acknowledgments}
We are grateful to our students Francesco Balassone and Ferdinando Simone D'Agostino for their help in the early stage of this work.

\section*{Data availability}
The tool is available and the files to reproduce our experiments are publicly available on the following URL: 
\href{https://github.com/dessertlab/DeVAIC}{https://github.com/dessertlab/DeVAIC}.

\bibliographystyle{plain}
\bibliography{mybibfile}

\end{document}